\documentclass[3p,times]{elsarticle}

\usepackage{ecrc}
\usepackage{epstopdf}

\volume{00}

\firstpage{1}

\journalname{Optics Communications}

\runauth{}

\jid{procs}

\usepackage{amssymb}
\usepackage{amsmath}

\usepackage[figuresright]{rotating}

\renewcommand{\eqref}[1]{Eq.(\ref{#1})}
\newcommand{\fig}[1]{Fig.~\ref{#1}}

\begin{document}

\begin{frontmatter}



\dochead{}

\title{Power Penalty Due to First-order PMD in Optical OFDM/QAM and FBMC/OQAM Transmission System}


\author[lab1]{Jianping Wang, Ke Zhang, Xianyu Du, He Zhen, Jing Yan}

\address[lab1]{Department of communication Engineering,30 Xueyuan Road, Haidian District, Beijing 100083 P. R.China}

\begin{abstract}

Polarization mode dispersion (PMD) is a challenge for high-data-rate optical-communication systems. More researches are desirable for impairments that is induced by PMD in high-speed optical orthogonal frequency division multiplexing (OFDM) transmission system. In this paper, an approximately analytical method for evaluating the power penalty due to first-order PMD in optical OFDM with quadrature amplitude modulation (OFDM/QAM) and filter bank based multi-carrier with offset quadrature amplitude modulation (FBMC/OQAM) transmission system is presented. The simulation results show that, compared with the single carrier with quadrature phase shift keying(SC-QPSK), both the OFDM/QAM and the FBMC/OQAM can decrease the power penalty caused by PMD by half. Furthermore, the FBMC/OQAM shows better power penalty immunity than the OFDM/QAM under the influence of first order PMD.

\end{abstract}


\end{frontmatter}



\section{Introduction}
\label{Sec1}

Optical fiber system has become a hot spot because of its ultra high speed, huge capacity and long haul transmission ability. Nowadays optical signal amplification and fiber dispersion compensation techniques are increasingly developed, and PMD has become the key limitation of transmission speed and distance\cite{boffi2008measurement,gene2010first,karlsson1998polarization,xu2010all}. PMD is a physical phenomenon caused by the birefringence of the optical fiber. When transmitting optical signal, PMD will cause different delays for different polarizations and the group delay difference between the slow and the fast modes is called differential group delay(DGD). When DGD is getting larger and can not be neglected compared with the signal bit duration, it will cause the pulse broadening and Inter-symbol Interference(ISI)\cite{boffi2008measurement}, then pulse distortion and system penalties occur. Different from the fiber chromatic dispersion(CD), PMD is a stochastic quantity influenced by the external conditions such as temperature or fiber vibration, et.al, which makes PMD particularly difficult to manage or compensate. In a specific system, dissipation due to PMD is determined by the theoretical structure model and physical factors (infrastructure, environmental factor, et.al), and different systems have different modulation techniques as well as PMD compensation techniques. 

Many researches in recent years are concentrated on PMD tolerance which based on different modulation modes and coding schemes\cite{liu2001influence,he2009performance}. In~\cite{liu2001influence}, the article proposed a 20Gb/s high-speed optical fiber transmission systems with Non Return to Zero (NRZ) and Return to Zero (RZ) code and the differences of the PMD-induced fiber channel is discussed by numerical simulation. In~\cite{he2009performance}, compared with on-off keying(OOK), it is demonstrated that differential phase shift keying(DPSK) signal has large first-order PMD tolerate ability in a 40Gb/s optical fiber transmission system. Other researches are focused on the PMD compensation techniques\cite{buchali2004adaptive,xie2003comparison, sunnerud2002comparison}, and the compensation of system PMD is demonstrated experimentally by electronical and optical compensators. Adaptive optical PMD compensation, which based on feedback, courts a balance between speed and accuracy. Recently, investigations pay more attention to the devised electronic PMD compensation schemes, such as iterative decoding techniques, maximum likelihood sequence estimation (MLSE), and transversal digital filtering\cite{mantzoukis2010outage,agazzi2005maximum,jager2006performance}. In ~\cite{mantzoukis2010outage}, PMD equalizers based on constant modulus algorithm (CMA) is presented in coherent optical polarization-division-multiplexed (PDM) QPSK systems taking account of PMD effect.

Recently, OFDM has been recommended as an effective PMD-resilient modulation format for high speed optical fiber transmission systems\cite{lowery2006orthogonal,shieh2007pmd,djordjevic2007pmd}. OFDM employs multi-carrier transmission of orthogonal, and has lower data rate subcarriers, therefore it simplifies PMD equalizer structure, and achieves high spectral efficiency in frequency-selective channels via the fast Fourier transform(FFT)\cite{cvijetic2008system}. In~\cite{djordjevic2007pmd}, the possibility of PMD compensation in fiber-optic communication systems with direct detection using a simple channel estimation technique and low-density parity-check (LDPC)-coded OFDM is demonstrated. In ~\cite{cvijetic2008system}, it has presented that, if it is not necessary for RF guard bands, OFDM capacitates high speed transmission with PMD tolerance which is at least twice  greater than that of uncompensated OOK at a given bit rate, while in systems with RF guard bands, a PMD tolerance trade-off proportional to guard band size was shown, where in guard band and constellation sizes may be viewed as design parameters. In ~\cite{cvijetic2007adaptive}, it has shown that, without requiring any feedback, OFDM can mitigate pulse distortion caused by all-order PMD in long-haul optic fiber communication systems. Like the OFDM, FBMC is another multi-carrier technology with higher spectral efficiency and has been perceived as an alternative to OFDM in recent years. While compared with OFDM, FBMC has a larger PMD tolerance because of its large stop-band attenuation and the frequency selective fading channel. Due to the advanced digital modulation technique,  the FBMC technique is quite fit for the high speed optical fiber transmission systems. As a result, in optical FBMC communication system, more research should be done on the impairment caused by the PMD.

In this paper, we focus on the system power penalty due to first-order PMD in multi-carrier optical transmission system which use OFDM/QAM and FBMC/OQAM modulation  format respectively, by comparing it with single carrier QPSK modulation, the theoretical model of power penalty in multi-carrier optical system impacted by first-order PMD is proposed, and then numerical simulation verification are given. The rest of the paper is organized as follow. In Section \ref{Sec2}, A brief introduction to the basic theory of OFDM/QAM and FBMC/OQAM modulation  has been given. And in Section \ref{Sec3}, the PMD principles, as well as the derivation of power penalty in optical OFDM/QAM and FBMC/OQAM is proposed. In Section \ref{Sec4}, the derivation results verified by the simulation results is proposed and finally the conclusion is given in Section \ref{Sec5}.


\section{Optical OFDM/QAM and FBMC/OQAM System Model}
\label{Sec2}

OFDM/QAM and FBMC/OQAM are kinds of multi-carrier modulation technique that can modulate and demodulate signals in frequency-domain by Inverse Fast Fourier Transform/Fast Fourier Transform (IFFT/FFT). OFDM technique can reduce the effects of dispersion and ISI efficiently and now is considered to be an effective solution to high speed optical communication in the future. High speed optical OFDM transmission will also be influenced by PMD effect as it is in single carrier modulation. Optical OFDM transmission systems have better anti-PMD effect ability compared with single carrier system because of the OFDM principles that separate a high speed data stream into several orthogonal low speed stream.

As we known that OFDM/QAM (i.e. CP-OFDM) has been widely used and considered as the core technique solution for next generation wireless communication, and FBMC/OQAM (i.e. OFDM/IOTA) is an alternative approach according to 3GPP protocols. Compared with traditional OFDM/QAM based on cyclic prefix (CP), FBMC/OQAM  without CP can achieve greater spectral efficiency, furthermore, FBMC/OQAM has better performance in wireless channel via choosing well time-frequency localized pulse shaping prototype filters\cite{du2007classic}. Applying OFDM techniques into high-capacity and high-speed optical fiber communication systems is a major research direction\cite{hussin2011performance,guoying2013asurvey}, and it will achieve a high flexibility and capacity in dynamic resource allocation and user access by combining with new technologies like PON, et.al\cite{wei2013design}. The basic principles of these two OFDM techniques are introduced bellow:

\subsection{OFDM/QAM System Model}

High speed information bit stream with bite rate $R_b=1/T_b$ is modulated in baseband using M-QAM modulation with symbol duration $T_s=T_b\log_2M$, and then divided in to $N$ parallel symbol streams which are filtered by a pulse shape function $g_{n,k}(t)$, the time-domain OFDM/QAM signal can be written in the following analytic form\cite{du2007classic}
\begin{equation}
s_{\textrm{QAM}}(t)=\sum_{k=1}^{+\infty}\sum_{n=1}^{N}a_{n,k}g_{n,k}(t)
\end{equation}
where
\begin{equation}
g_{n,k}(t)=e^{j2\pi nFt}g(t-kT)
\end{equation}
$F$ denotes the inter-carrier frequency spacing and $T$ is the OFDM symbol duration. $a_{n,k}$ represents the QAM baseband modulation output data on the $n$th subcarrier at time index $k$. In a OFDM/QAM system, $F=1/NT_s=\nu_0$, $T=\tau_0$ and in order to satisfy the orthogonality, $\tau_0\nu_0=1$, and the prototype function is defined as 
\begin{equation}
g(t)=\left\{\begin{array}{lc}
1/\sqrt{\tau_0},  &  0\leq t < \tau_0  \\
0,           &  \textrm{elsewhere}
\end{array}\right.
\end{equation}

\subsection{FBMC/OQAM System Model}

Under the same initial conditions as OFDM/QAM, the time domain FBMC/OQAM signal can be expressed as \eqref{OQAMtimedomain}\cite{du2007classic,siohan2002analysis}
\begin{equation}\label{OQAMtimedomain}
s_{\textrm{OQAM}}(t)=\sum_{k=1}^{+\infty}\sum_{n=1}^{N}a_{n,k}g_{n,k}(t)
\end{equation}
where 
\begin{equation}
g_{n,k}(t)=e^{j2\pi n\nu_0t}g(t-k\tau_0)\times e^{j(n+k)\pi/2},\:\nu_0\tau_0=1/2
\end{equation}
$g(t)$ is the prototype pulse shaping function that can be different from rectangular window, for example, Extended Gaussian Function (EGF) and Isotropic Orthogonal Transform Algorithm (IOTA) Function, et.al. Unlike the original OFDM/QAM, FBMC/OQAM employs a modified inner product by taking a real component to maintain the orthogonality among the synthesis and analysis basis, as show in bellow
\begin{equation}\label{modifiedinnerproduct}
\mathcal{R}\left\{g_{n,k}^*\times g_{n',k'}\right\}=\left\{\begin{array}{lc}
1,  &   (n,k)=(n',k') \\
0,  &   (n,k)\neq(n',k')
\end{array}\right.
\end{equation}

The purpose of pulse shaping in FBMC/OQAM is to find an efficient transmitter and a corresponding receiver waveform for the current channel condition\cite{schafhuber2002pulseshaping,bass2004pulseshaping}, a well time-frequency localized waveform should satisfy
\begin{equation}
\frac{\tau_0}{\Delta t} = \frac{\nu_0}{\Delta f}
\end{equation}
where $\Delta t$ and $\Delta f$ is the RMS delay spread and Doppler spread, respectively.


\section{Power Penalty duo to First-order PMD}
\label{Sec3}

Currently, OFDM is proposed to be a promising modulation technique for high-speed optical transmission systems,  owing to high tolerance to CD and PMD. However, PMD still degrades the performance of the high-speed optical transmisson systems, due to lacking of mature compensation techniques. Subsequent bit error and system power penalty analysis seeks to assess in order to evaluate the PMD tolerance of high speed fiber OFDM/QAM and FBMC/OQAM transmission system.

\begin{figure}[!t]
\begin{center}
\includegraphics[width=0.5\linewidth]{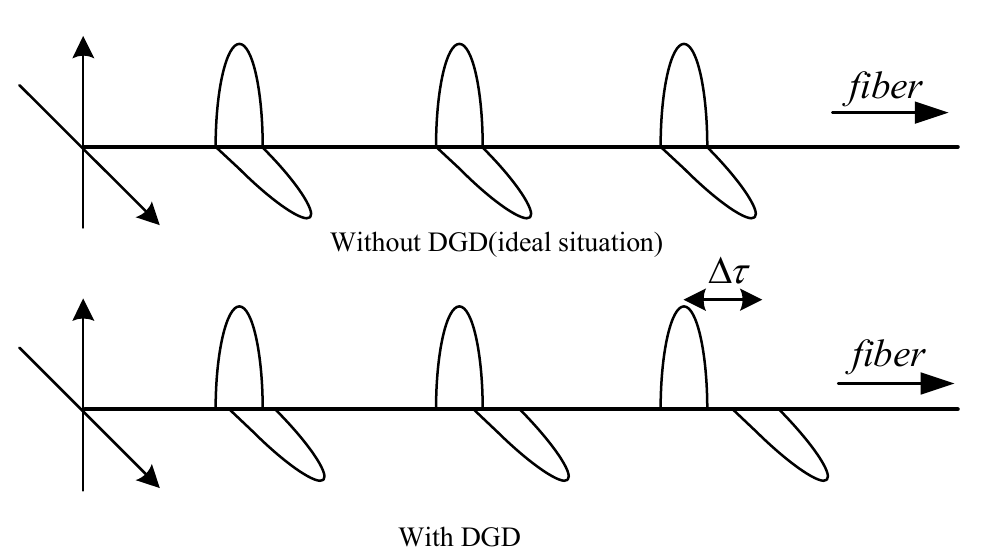}
\caption{Time-domain behavior of PMD in a short birefringent fiber.}\label{DGDfig}
\end{center}
\end{figure}

In single mode fiber (SMF) transmission, optical signals are composed by two orthogonally polarized $HE_{11}$ mode. If the SMF is ideal, the two polarized mode have the same refractive index and transmitting speed, so there won't be any PMD as it shown in \fig{DGDfig}. However, in practical fiber, it's impossible to achieve identical refractive index, thus there will be a different delay between the two polarized mode and causes the DGD, as shown in \fig{DGDfig}, this phenomenon is called PMD. In long haul and high speed optical fiber communication system, pulse broadening caused by PMD effect can lead to serious ISI, which will degrade the system performance, and that is why PMD has been considered as a key factor after CD and fiber attenuation.

\begin{figure}[!t]
\begin{center}
\includegraphics[width=0.7\linewidth]{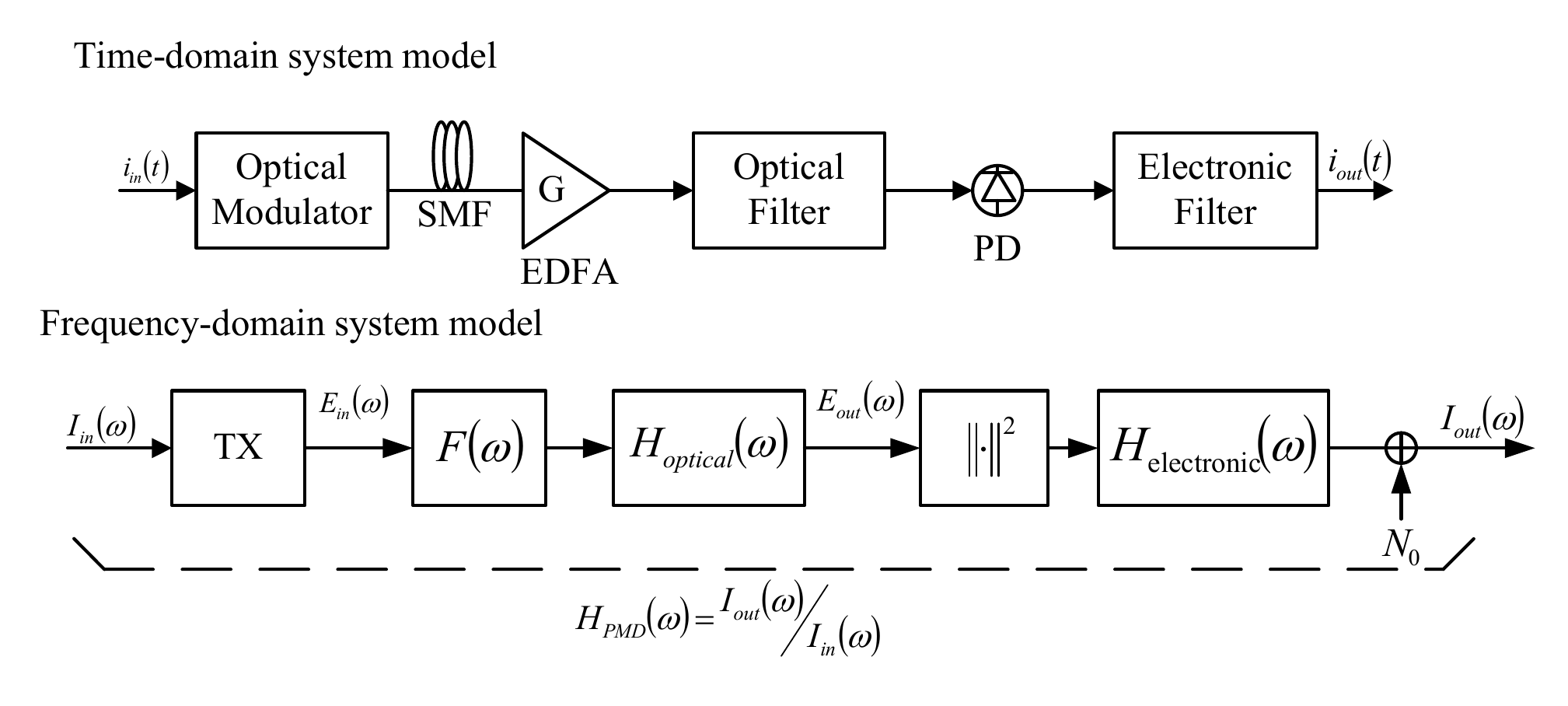}
\caption{System model and corresponding lowpass equivalent.}\label{blockfig}
\end{center}
\end{figure}

\fig{blockfig} illustrate the block diagram of basic optical fiber transmission system and it's frequency-domain equivalent, the input electronic signal after electro-optical modulation (EOM) is set as $i_{in}(t)$, the envelope of the input signal is  $\tilde{x}(t)=\alpha i_{in}(t)$ and $\alpha$ is a proportionality coefficient. The Jones vector of the resulting electronic field at the fiber input is given by $\tilde{E}_{in}(t)$ as\cite{cvijetic2008system}
\begin{equation}
\tilde{E}_{in}(t)=\tilde{E}_{in}(t)\hat{e}_{in}=R[\sqrt{\tilde{x}(t)}e^{j\omega_0t}]\hat{e}_{in}
\end{equation}
where $\hat{e}_{in}=[\hat{e}_{1},\hat{e}_{2}]^T$ is the polarization state Jones vector, whose entries denote the two orthogonal Principle State of Polarizations (PSPs) at the fiber input. Under the First-order PMD approximation, PMD vector $\vec{\Omega}=\vec{\Omega}_0=\Delta\tau\hat{e}_1$, $\hat{e}_1$ is the unit vector in fast PSP. The Fourier transform of fiber output signal $\hat{E}_{out}(t)$ is given by\cite{cvijetic2008system}
\begin{equation}
\hat{E}_{out}(\omega)=\textbf{F}(\omega)\hat{E}_{in}(\omega)=\textbf{RU}(\omega)\textbf{R}^{-1}\hat{E}_{in}(\omega)
\end{equation}
where $\hat{E}_{in}(\omega)=\mathcal{F}(\tilde{E}_{in}(t))$, $\textbf{F}(\omega)=\textbf{RU}(\omega)\textbf{R}^{-1}$ denotes the fiber Jones matrix. $\textbf{R}$ denotes the random, frequency-independent rotation matrix and $\textbf{U}(\omega)$ denotes the time delay matrix caused by first-order PMD which can be expressed as \cite{forestieri2004exact}:
\begin{equation}
\textbf{U}(\omega)=\left[\begin{array}{cc}
e^{j\omega\Delta\tau/2}  &  0   \\
0            &  e^{-j\omega\Delta\tau/2} 
\end{array}\right]
\end{equation}
where $\Delta \tau$ is the differential group delay (DGD). And the explicit expression for $\textbf{R}$ is given by\cite{forestieri2004exact}
\begin{equation}
\textbf{R}=\left[\begin{array}{cc}
r_{1}  & -r_{2}^* \\
r_{2}  & r_{1}^*
\end{array}\right]
\end{equation}
where
\begin{equation}\begin{split}
r_{1}= & \cos{\theta}\cos{\phi}-j\sin{\theta}\sin{\phi} \\
r_{2}= & \sin{\theta}\cos{\phi}+j\cos{\theta}\sin{\phi}
\end{split}\end{equation}
where $\theta$,$\phi$ are independent random variables,representing the fast PSP azimuth and ellipticity angle respectively.

And then we can get\cite{cvijetic2008system}
\begin{equation}
\hat{E}_{out}(t)=c_1E_{in}(t+\Delta\tau/2)\hat{e}_1+c_2E_{in}(t-\Delta\tau/2)\hat{e}_2
\end{equation}
where $c_1,c_2$ depend on the rotation matrix $\textbf{R}$, and can be expressed by
\begin{equation}\begin{split}
c_{1}= & |\cos{\varphi}| \\
c_{2}= & |\sin{\varphi}|
\end{split}\end{equation}
where $2\varphi$ denotes the angle between the fast PSP and the signal polarization state in Stokes space.After photoelectric detection (PD) and post-detection filtering, the electronic signal\cite{cvijetic2008system}
\begin{equation}\begin{split}\label{detection}
i_{out}(t) & = \rho|\hat{E}_{out}(t)|^2  \\
           & = \rho\left[|c_1\sqrt{\tilde{x}(t+\Delta\tau/2)}|^2+|c_2\sqrt{\tilde{x}(t-\Delta\tau/2)}|^2\right]  \\
           & = \rho\alpha\left[\gamma i_{in}(t+\Delta\tau/2)+(1-\gamma)i_{in}(t-\Delta\tau/2)\right]
\end{split}\end{equation}
where $\rho$ is photoelectric detector sensitivity and $\gamma=|c_1|^2$ is the PSP power splitting ratio.

In \eqref{detection}, it is obvious that first-order PMD causes pulse broadening and results in adjacent pulses overlap, which will finally cause the Power Penalty at the receiver. System Power Penalty due to PMD effect is defined as the difference of receiver sensitivity (in dB) between two conditions, with or without PMD effect. Considering first-order PMD approximation, the input optical signal is only divided into two orthogonal polarized mode and causes one DGD $\Delta\tau$, and leads to power penalty\cite{poole1991fading}.

Under this assumption, we assume that the received signal $i_{out}(t)$ can be determined by input signal $i_{in}(t)$ after fiber transmission via \eqref{detection}
\begin{equation}
i_{out}(t)=\gamma i_{in}(t+\Delta\tau/2)+(1-\gamma)i_{in}(t-\Delta\tau/2)
\end{equation} 
The Root-Mean-Square (RMS) pulse width $\delta_2$ of the output signal is given by\cite{poole1991fading}
\begin{equation}\label{delta2}
\delta_2^2=\frac{\int_{-\infty}^{+\infty}t^2i_{out}(t)dt}{\int_{-\infty}^{+\infty}i_{out}(t)dt}-[\frac{\int_{-\infty}^{+\infty}ti_{out}(t)dt}{\int_{-\infty}^{+\infty}i_{out}(t)dt}]^2
\end{equation}

Assuming that $i_{in}(t)$ is symmetrical about $t=0$, therefore $i_{in}(t+\Delta\tau/2)$ and $i_{in}(t-\Delta\tau/2)$ are symmetrical about $t=0$ too, so \eqref{delta2} can be expressed as
\begin{equation}
\delta_2^2=\frac{\int_{-\infty}^{+\infty}t^2i_{in}(t)dt}{\int_{-\infty}^{+\infty}i_{in}(t)dt}+\Delta\tau^2\gamma(1-\gamma)
\end{equation}
When $\Delta\tau=0$ (without PMD), RMS pulse width $\delta_1$ of the input signal is given by
\begin{equation}\begin{split}\label{delta1}
\delta_1^2 & = \frac{\int_{-\infty}^{+\infty}t^2i_{in}(t)dt}{\int_{-\infty}^{+\infty}i_{in}(t)dt}-[\frac{\int_{-\infty}^{+\infty}ti_{in}(t)dt}{\int_{-\infty}^{+\infty}i_{in}(t)dt}]^2 \\
           & = \frac{\int_{-\infty}^{+\infty}t^2i_{in}(t)dt}{\int_{-\infty}^{+\infty}i_{in}(t)dt}
\end{split}\end{equation}
Furthermore, we can observe that\cite{poole1991fading}
\begin{equation}
\delta_2^2=\delta_1^2+\Delta\tau^2\gamma(1-\gamma)
\end{equation}
Irrespective of the polarization dependent loss (PDL), power penalty due to PMD effect $\epsilon$ can be represented as
\begin{equation}
\epsilon(dB)=10lg\frac{\delta_2}{\delta_1}=5lg(1+\frac{\Delta\tau^2\gamma(1-\gamma)}{\delta_1^2})
\end{equation}
where $\Delta\tau$ is very small in a general way, so 
\begin{equation}
\epsilon(dB)\approx5\frac{\Delta\tau^2\gamma(1-\gamma)}{\delta_1^2}
\end{equation}

For general system, the RMS pulse width $\delta_1$ of input signal is proportional to bit interval $T_b$, power penalty $\epsilon$ can be represented as\cite{poole1991fading}
\begin{equation}\label{penalty}
\epsilon=\frac{A}{T_b^2}\Delta\tau^2\gamma(1-\gamma)
\end{equation}
where $A$ is a coefficient concerned with pulse shape, modulation format and receiver characters, et.al. $T_b$ is bit interval and $\Delta\tau$ is the instant DGD. The PSP power splitting ratio $0<\gamma<1$. For a fixed DGD, power penalty is maximized when $\gamma=0.5$, so it should be considered as a requirement while designing a system to ensure performance. In the follow simulations, $\gamma$ is fixed to be 0.5.

In original single carrier optical fiber communication systems, power penalty due to PMD effect $\epsilon$ (in dB) can be seen in \eqref{penalty} , and when it comes to the multi-carrier optical fiber transmission, power penalty due to PMD can be deduced by the following content.

Assuming an OFDM/QAM system with symbol duration $\tau_0=T$, sub-carrier number $N$ and inter-carrier frequency spacing $\nu_0=F=1/T$. $\Delta\tau$ represents the DGD caused by first-order PMD. The baseband modulation scheme is SC-QPSK and the OFDM signal bandwidth $BW=N\times F$, bit rate $R_b=N\times F\times \log_2 M$ (for QPSK, 4QAM, $M=4$ and for 16QAM, $M=16$, et.al) and bit duration $T_b=1/R_b$.

For FBMC/OQAM, inter-carrier frequency spacing $\nu_0=F=1/T$ and $\tau_0=T/2$. Firstly, we can get the power penalty $\epsilon_n$ in $n$th sub-carrier via its angular frequency $\omega_n$
\begin{equation}
\omega_n=2\pi\nu_0 n
\end{equation}

For OFDM/QAM and FBMC/OQAM, the cycle of the $n$th sub-carrier $T_n=1/\nu_0 n=T/n$ and put it into \eqref{penalty}
\begin{equation}
\epsilon_n=\frac{A}{T_n^2}\Delta\tau^2\gamma(1-\gamma)=\frac{A\gamma(1-\gamma)n^2\Delta\tau^2}{T^2}
\end{equation}

Assuming that transmitting power of each sub-carrier is $P_0$ without PMD can satisfy the requirement of receiver PMD while the $n$th sub-carrier transmitting power is $P_n$ with first-order PMD to achieve the same performance. According to the definition of power penalty we can define the power penalty of the $n$th sub-carrier as
\begin{equation}\label{epsilon_n}
\epsilon_n=10\log\frac{P_n}{P_0}
\end{equation}
then the total power penalty of the multi-carrier system is given by
\begin{equation}\label{OFDMpenalty}
\epsilon=10\log\frac{\sum_{n=1}^{N}P_n}{NP_0}
\end{equation}
put \eqref{epsilon_n} into \eqref{OFDMpenalty}
\begin{equation}\label{OFDMpenalty2}
\epsilon=10\log\left(\frac{1}{N}\sum_{n=1}^{N}10^{\frac{\epsilon_n}{10}}\right)
\end{equation}

Equation~\eqref{OFDMpenalty} is the conclusion that the theoretical expression of power penalty due to first-order PMD in OFDM/QAM and FBMC/OQAM system and in the next section, numerical simulation will be given to prove its correctness.

\section{Numerical Simulation}
\label{Sec4}

In this section,focused on studying the power penalty due to first-order PMD in SC-QPSK, OFDM/QAM and FBMC/OQAM system respectively. Specially, in OFDM/QAM and FBMC/OQAM simulation, we fixed inter-carrier frequency spacing $\nu_0=100$MHz and PSP power splitting ratio $\gamma=0.5$, sub-carrier number $N=64$ and $N=128$ respectively. The prototype pulse shaping function in FBMC/OQAM is set to be Square Root Raised Cosine (SRRC) filter with the length of $L=4N$. Obviously, in order to compare with the multi-carrier condition on the PMD tolerance problem, we set the SC-QPSK modulation with the same transmission bit rate as the OFDM/QAM and FBMC/OQAM. For example, SC-QPSK modulation bit rate $R_b=2N\nu_0=25.6$Gb/s for $N=128$ and $R_b=2N\nu_0=12.8$Gb/s for $N=64$. Other factors (like FEC et.al) remain unchanged when comparing OFDM/QAM with FBMC/OQAM.

\begin{figure}[!t]
\begin{center}
\includegraphics[width=0.5\linewidth]{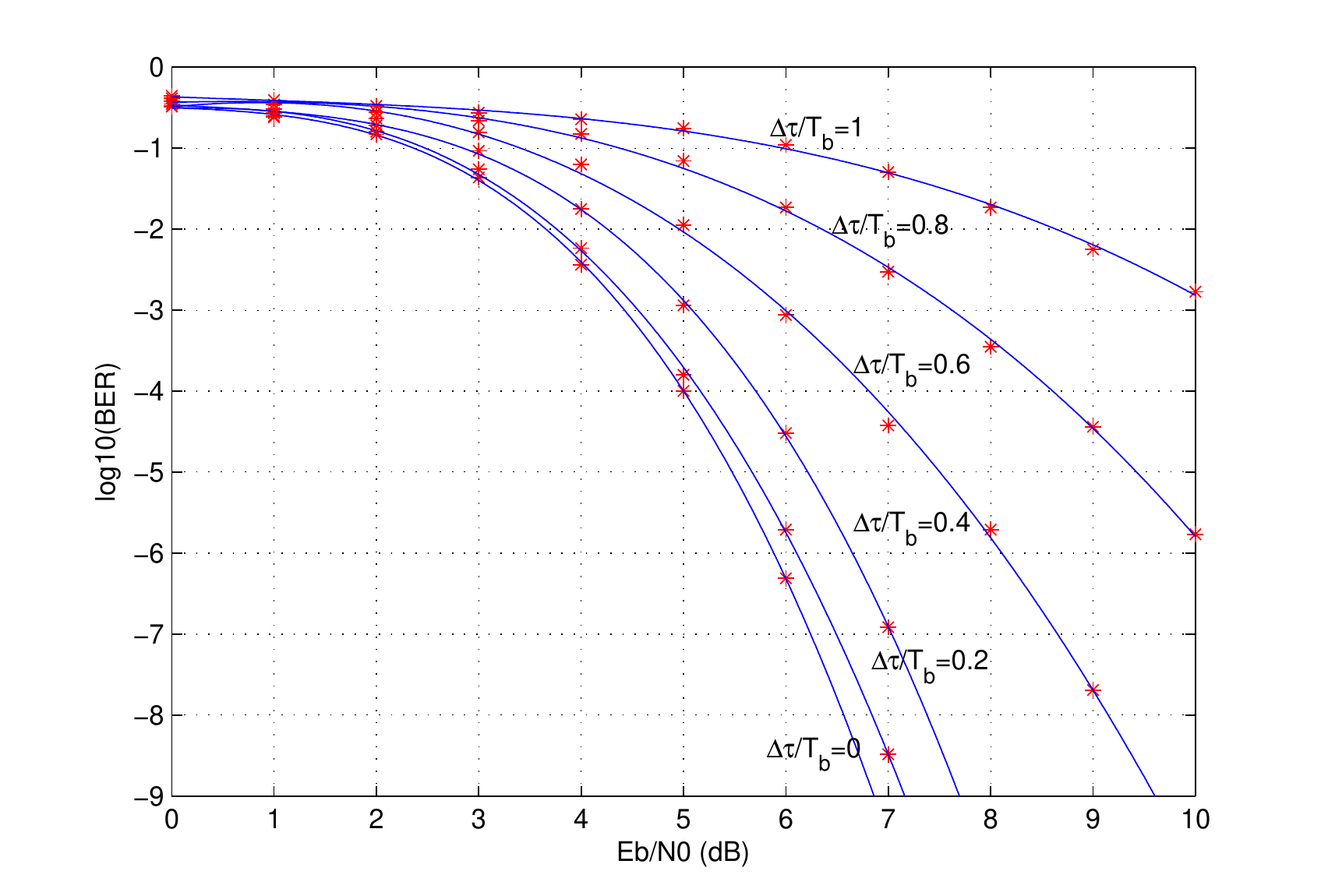}
\caption{Simulation BER versus $E_b/E_o$ results for several $\Delta\tau/T_b$ values under OFDM/QAM.}\label{OFDMBER}
\end{center}
\end{figure}

The Bit Error Rate (BER) vs Signal to Noise Ratio (SNR) simulation results of the OFDM/QAM system with $N=128$ and inter-carrier frequency spacing $\nu_0=100$MHz is shown in \fig{OFDMBER},  DGD with 0, 0.2, 0.4, 0.8 and 1 times of the bit duration $T_b$. caused by first-order PMD is separately simulated. It's clearly shown in the figure that system BER can reach $10^{-9}$ when $E_b/N_0$ is about 6.8dB without PMD(DGD$=0$) and with the growth of DGD, we have to increase the transmitter power to improve the channel SNR in order to maintain the system BER performance in $10^{-9}$. When DGD$=0.4\times /2N\nu_0=15.6ps$, thus 0.4 times of bit duration, the $E_b/N_0$ is 7.7dB at the point that BER is $10^{-9}$, and the power penalty due to PMD in this situation $\epsilon=7.8-7.1=0.7(dB)$. Comparing all the simulation results, We can find that the power penalty due to PMD effect is growing more faster when DGD is getting bigger, which means the signal distortion is getting more worse.

\begin{figure}[!t]
\begin{center}
\includegraphics[width=0.5\linewidth]{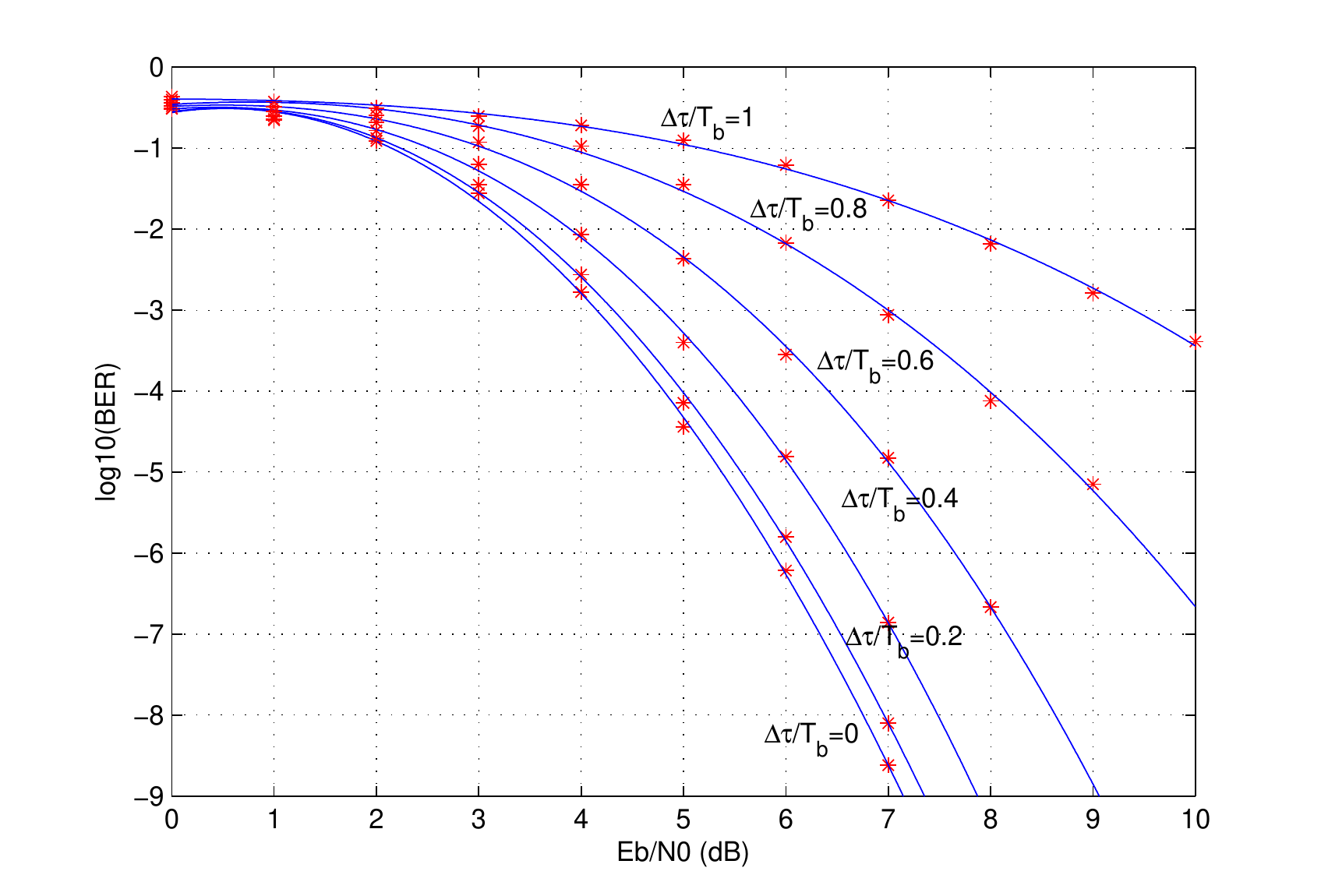}
\caption{Simulational BER versus $E_b/E_o$ results for several $\Delta\tau/T_b$ values under FBMC/OQAM.}\label{OQAMBER}
\end{center}
\end{figure}

\begin{figure}[!t]
\begin{center}
\includegraphics[width=0.5\linewidth]{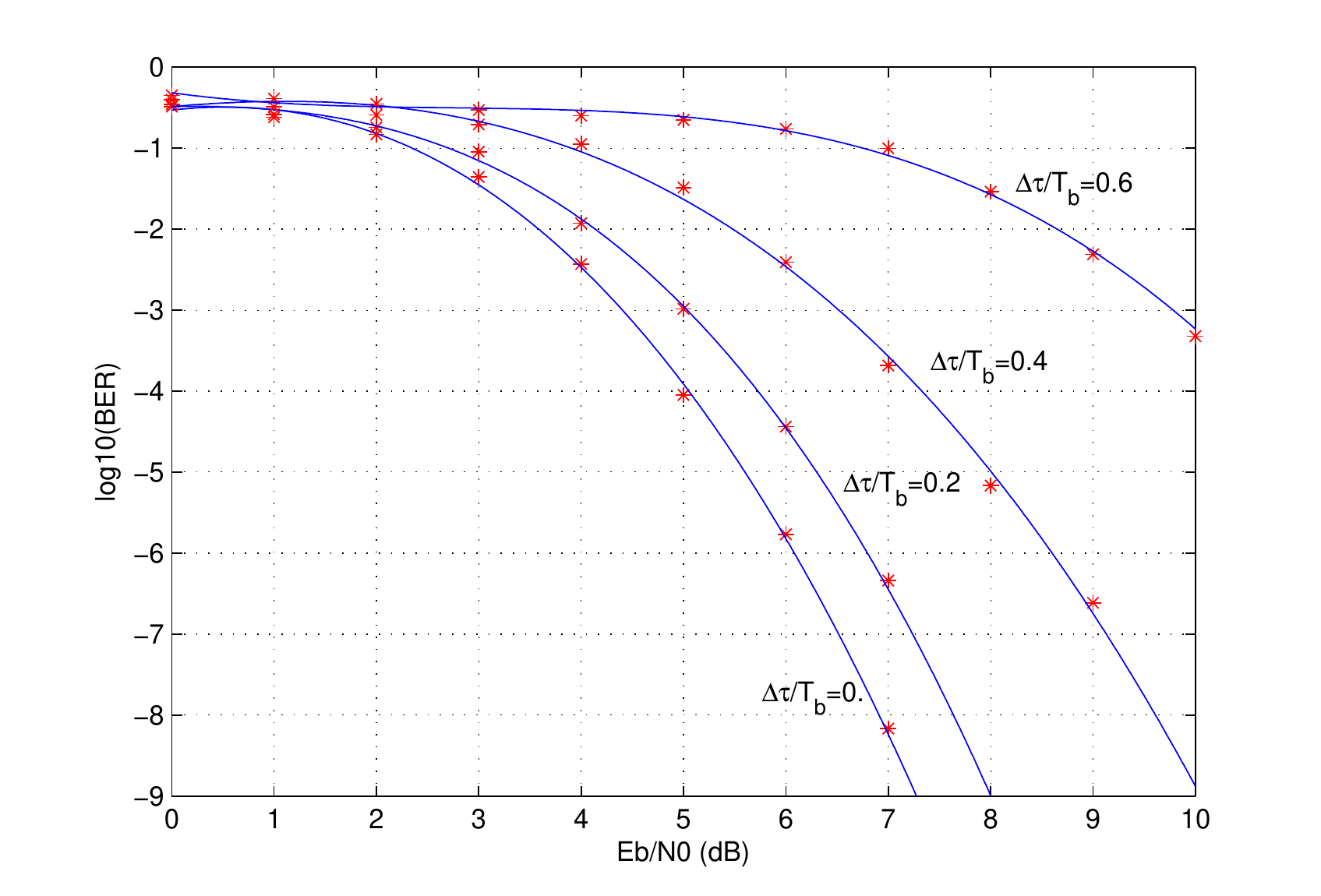}
\caption{Simulation BER versus $E_b/E_o$ results for several $\Delta\tau/T_b$ values under SC-QPSk.}\label{QPSKBER}
\end{center}
\end{figure}

 The BER performance of the FBMC/OQAM modulation scheme is show in \fig{OQAMBER}. The simulation condition is the same with that of the OFDM/QAM system Without the influence of PMD effect, the system BER reaches $10^{-9}$ at $E_b/N_0=$7.1dB, and when DGD is 0.4 times of bit duration ,$E_b/N_0$ should be 7.8dB to achieve the same BER performance. The power penalty $\epsilon=7.8-7.1=0.7$dB at that moment. Same with the OFDM/QAM the power penalty in the FBMC/OQAM  caused by the first-order PMD is growing faster when DGD is getting bigger and leads to more serious signal distortion.

The difference of PMD tolerance between single carrier and multi carrier system can be seen from \fig{QPSKBER} which demonstrate the BER performance of a 25.6Gb/s SC-QPSK signal with different DGD due to first-order PMD . Same as above, $E_b/N_0$ is 7.2dB and 10.1dB when DGD is 0 and $0.4T_b$ respectively, and derivatives the power penalty is about 2.9dB. By observing \fig{OFDMBER}, \fig{OQAMBER} and \fig{QPSKBER}, we can find that both OFDM/QAM and FBMC/OQAM has better anti-PMD ability than SC-QPSK.
\begin{figure}[!t]
\begin{center}
\includegraphics[width=0.5\linewidth]{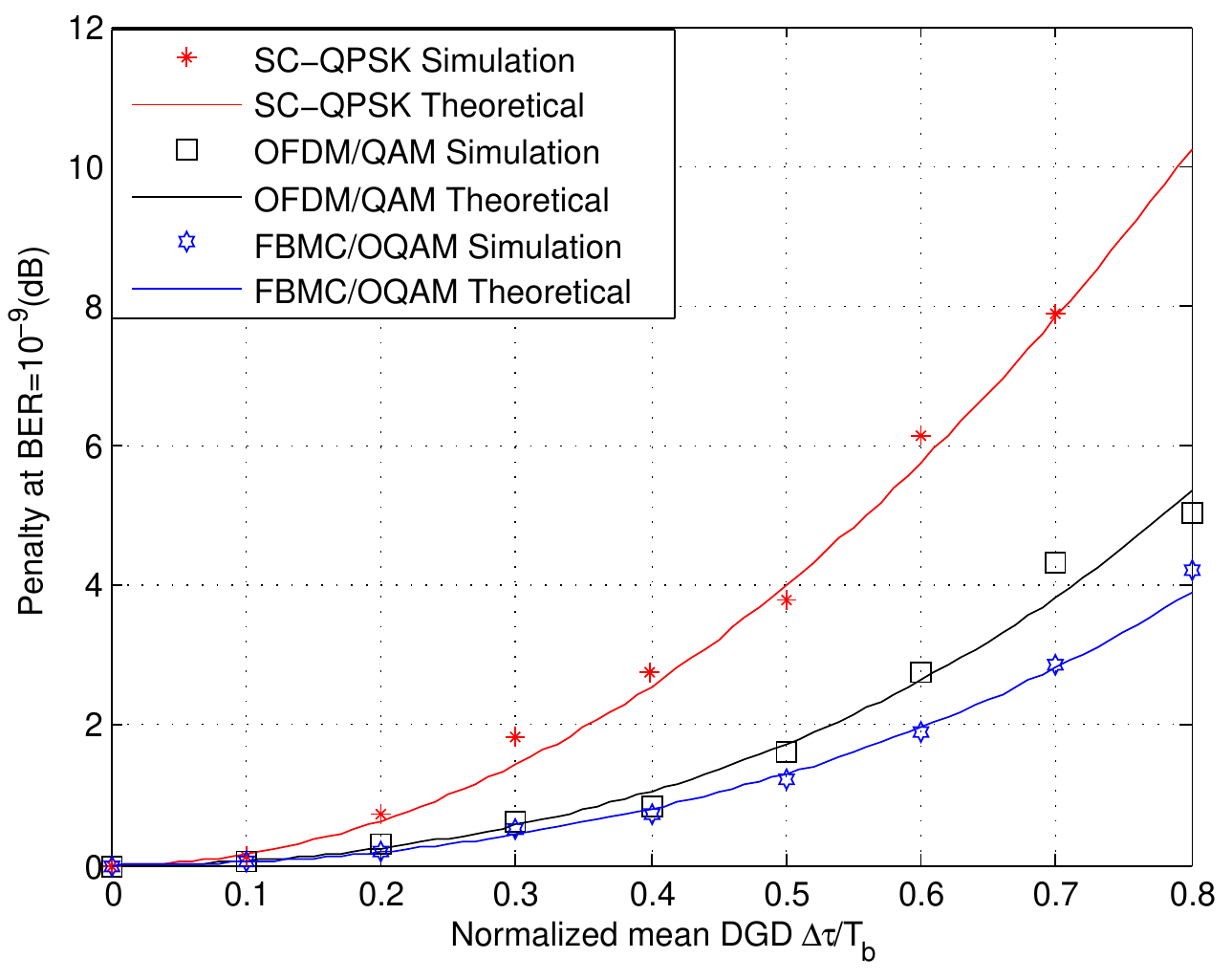}
\caption{Power penalty under SC-QPSK,OFDM/QAM and FBMC/OQAM for N=128.}\label{PMDpp128}
\end{center}
\end{figure}

\begin{figure}[!t]
\begin{center}
\includegraphics[width=0.5\linewidth]{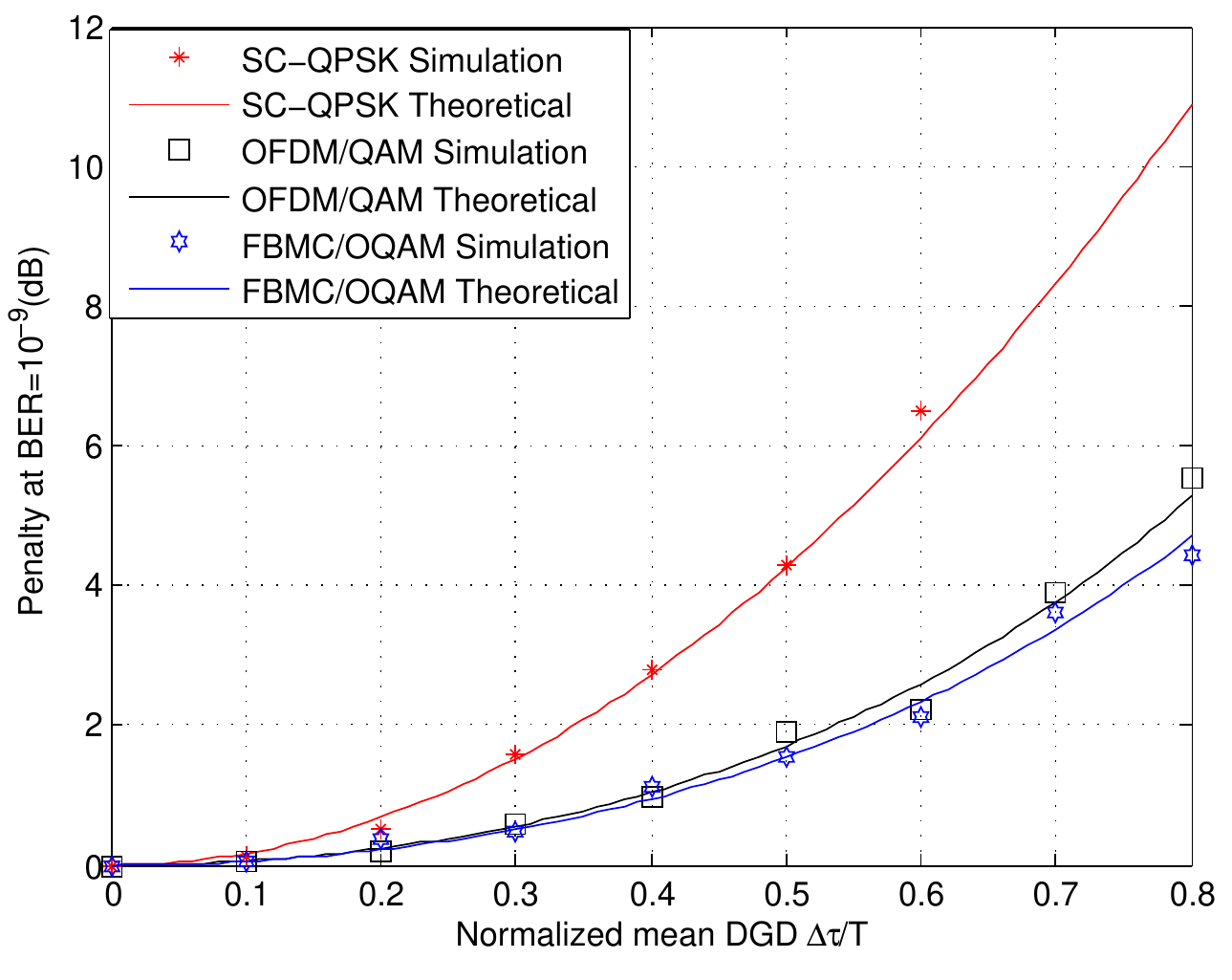}
\caption{Power penalty under SC-QPSK,OFDM/QAM and FBMC/OQAM for N=64.}\label{PMDpp64}
\end{center}
\end{figure}

\fig{PMDpp128} presents the power penalty with different normalized mean DGD with the SC-QPSK, OFDM/QAM and FBMC/OQAM modulation at the same transmission bit rate of 25.6Gb/s, and the sub-carrier number $N=128$. Correctness can be verified by the simulation results that the SC-QPSK power penalty matches well with the theoretical curve which is given by \eqref{penalty}. Theoretical curves of OFDM/QAM and FBMC/OQAM power penalty due to first-order PMD which are given by our derivation in \eqref{OFDMpenalty2} are also shown in this figure , and it's clear to see that the simulation results are consistent with theoretical values pretty well. The coefficient $A$ is set as 68, 64 and 60 in SC-QPSK, OFDM/QAM and FBMC/OQAM respectively as its value is dependent on multi-factors like modulation schemes, pulse shaping techniques and receive modes, et.al, that is, $A$ is a different value for different systems.
 
When power penalty is 1dB for SC-QPSK, OFDM/QAM and FBMC/OQAM systems, $\Delta\tau/T_b$ is about 0.24, 0.41 and 0.45 in \fig{PMDpp128}, and OFDM/QAM and FBMC/OQAM have about twice more PMD tolerance than SC-QPSK from this view point and show better anti-PMD abilities. FBMC/OQAM shows better performance than OFDM/QAM at the same time because of its out of band attenuation. For example, the power penalty of OFDM/QAM and FBMC/OQAM is 0.9dB and 0.7dB respectively when $\Delta\tau/T_b=0.4$, thus the latter one is 0.2dB less than the former. And when $\Delta\tau/T_b=0.6$, the gap between these two schemes is increased to about 1dB. We can deduce a conclusion that FBMC/OQAM will achieve better anti-PMD ability with the growth of PMD effect.

\fig{PMDpp64} presents the same situation with sub-carrier number $N=64$, and the same conclusion can be deduced from this figure compared with \fig{PMDpp128}. comparing these two figure we can find that the difference between OFDM/QAM and FBMC/OQAM is turning smaller with a lower number of sub carrier which give us the conclusion that FBMC/OQAM shows better anti-PMD performance in power penalty than OFDM/QAM with larger sub-carrier number $N$.

\section{Conclusion}
\label{Sec5}

In this paper, we discussed the System Power Penalty due to first-order PMD effect in multi-carrier optical communication system, especially two modulation schemes OFDM/QAM and FBMC/OQAM. Theoretical derivation of the multi-carrier condition were given by comparing with original single carrier situation at the first, and then confirmed its validity via the numerical simulation. Through the simulation results we can find that the power penalty due to first-order PMD in OFDM/QAM and FBMC/OQAM systems is about a half smaller than that of the single carrier SC-QPSK system at the same transmitting bit rate, and with the growth of sub-carrier number and bit rate, the latter one can achieve better PMD resistance ability than the the former.

\section*{Acknowledgments}

This research is supported by the Fundamental Research Funds for the Central Universities (No.FRF-TP-09-015A), and also supported by the National Natural Science Foundation of P.R.China (No.61272507).





\bibliographystyle{elsarticle-num}
\bibliography{IEEEabrv,ke}







\end{document}